\documentclass[aps, prl, reprint,floatfix,superscriptaddress]{revtex4-1}
\usepackage[utf8]{inputenc}
\usepackage{amsmath}
\usepackage{dcolumn}
\usepackage{natbib}
\usepackage{soul}
\usepackage{ulem}
\usepackage{graphicx}
\usepackage{mathrsfs}
\usepackage{bm}
\makeatletter
\newcommand{\RNum}[1]{\uppercase\expandafter{\romannumeral #1\relax}}
\usepackage{float}
\makeatother
\usepackage{color,soul}
\usepackage{amssymb}
\usepackage{gensymb}
\usepackage{xcolor}

\begin{document}

\author{Yifan Zhou}
\affiliation{NanoSpin, Department of Applied Physics, Aalto University School of Science, P.O. Box 15100, FI-00076 Aalto, Finland}
\author{Rhodri Mansell}
\email{rhodri.mansell@aalto.fi}
\affiliation{NanoSpin, Department of Applied Physics, Aalto University School of Science, P.O. Box 15100, FI-00076 Aalto, Finland}
\author{Tapio Ala-Nissila}
\affiliation{QTF Center of Excellence, Department of Applied Physics, Aalto University, FI-00076 Aalto, Espoo, Finland}
\affiliation{Interdisciplinary Centre for Mathematical Modelling and Department of Mathematical Sciences, Loughborough University, Loughborough, Leicestershire LE11 3TU, UK}
\author{Sebastiaan van Dijken}
\affiliation{NanoSpin, Department of Applied Physics, Aalto University School of Science, P.O. Box 15100, FI-00076 Aalto, Finland}

\title{Thermal motion of skyrmion arrays in granular films}

\begin{abstract}
Magnetic skyrmions are topologically-distinct swirls of magnetic moments which display particle-like behaviour, including the ability to undergo thermally-driven diffusion. In this paper we study the thermally activated motion of arrays of skyrmions using temperature dependent micromagnetic simulations where the skyrmions form spontaneously. In particular, we study the interaction of skyrmions with grain boundaries, which are a typical feature of sputtered ultrathin films used in experimental devices.  We find the interactions lead to two distinct regimes. For longer lag times the grains lead to a reduction in the diffusion coefficient, which is strongest for grain sizes similar to the skyrmion diameter. At shorter lag times the presence of grains enhances the effective diffusion coefficient due to the gyrotropic motion of the skyrmions induced by their interactions with grain boundaries. For grain sizes significantly larger than the skyrmion diameter clustering of the skyrmions occurs in grains with lower magnetic anisotropy. 

\end{abstract}

\maketitle

The single and collective diffusion of particles has been investigated in a wide variety of systems using a variety of theoretical models \cite{Ala-NissilaDIffusionReview,MetzlerDiffusionReview}. Magnetic skyrmions, which are topologically protected quasi-particles, also show random motion under thermal fields, which has been investigated theoretically \cite{Schutte2014,Miltat2018}, by numerical methods \cite{Troncoso2014,Rohart2016,rozsa2017temperature} as well as through experimental observations \cite{Nozaki2019,ZazvorkaSkyrmionDiffusion,Zhao2020}. It has been suggested that the random motion of skyrmions can be utilised in probabilistic computing \cite{Pinna2018,BourianoffReservoirComputing,ZazvorkaSkyrmionDiffusion}, with the possibility of controlling the diffusion coefficient by electric field \cite{Nozaki2019}. However, in order to create such devices the realities of experimental systems need to be taken into account. Particularly, experimental systems contain pinning sites, which have been shown to play an important role in skyrmion dynamics in various simulations of single skyrmion systems under temperature \cite{Miltat2018,Juge2018,Uzdin2018} and activation of skyrmions by electric currents \cite{JVKim2017,Salimath2019,Gong2020,IwasakiSkUniversal2013,LegrandSub100Motion2017}. Furthermore, by assuming skyrmions are rigid particles, molecular dynamics simulations have also been used to study the properties of skyrmion lattices pinned by point defects \cite{Brown2019} and the corresponding depinning transition under a driving force \cite{Xiong2019,Reichhardt2015}. However, despite the large number of papers on the interaction of skyrmions with pinning sites \cite{2021arXiv210210464R}, there is still a need to better understand the thermal motion of skyrmions by combining the multi-skyrmion approach envisaged in devices, with modelling that reflects the properties of the experimental films. Single skyrmion approaches cannot account for the collective effects in skyrmion arrays, while the rigid particle assumption typically used in molecular dynamics does not reflect the deformations of skyrmions which occur in real systems. 

Here, we study the thermally-driven motion of skyrmions in simulations which emulate sputtered ultrathin multilayer heterostructures, a technologically relevant skyrmion system \cite{Fert2013Skyrmiontrack,jiang2015blowing,ZhangSkyrmionLogic}. In particular these layers have a grain structure, with variation of the underlying magnetic properties occurring between neighbouring grains \cite{JVKim2017}. One convenient way of understanding the properties of skyrmions under external stimulus is through the Thiele equation, which assumes a rigid skyrmion shape \cite{thiele1973steady}. 
Skyrmion motion can then be approximated by defining $\bm{m}(\bm{r},t)=\bm{m}(\bm{r}-\bm{C}(t))$, where $\bm{m}$ is the magnetic moment, $\bm{r}$ is the position vector of the spins relative to the skyrmion center, and $\bm{C}(t)$ is the centre-of-mass coordinate of the skyrmion. The dynamics of a rigid skyrmion are then given by:

\begin{equation}\label{Thiele}
M_{s}{\gamma}^{-1}\bm{G}\times \dot{\bm{C}}+M_{s}{\gamma}^{-1}\alpha {\mathcal{D}} \dot{\bm{C}}=\bm{F},
\end{equation}
where
\begin{equation}\label{G}
 \begin{split}
 &G=\iint \mathrm{d}x \mathrm{d}y \bm{m} \cdot (\frac{\partial \bm{m}}{\partial x} \times \frac{\partial \bm{m}}{\partial y}),
\\&\mathcal{D}=\iint \mathrm{d}x \mathrm{d}y (\frac{\partial \bm{m}}{\partial x} \cdot \frac{\partial \bm{m}}{\partial x} + \frac{\partial \bm{m}}{\partial y} \cdot \frac{\partial \bm{m}}{\partial y})/2,
\\&\bm{F}=-\nabla U,
 \end{split}
\end{equation}
$M_s$ is the saturation magnetization, $\gamma$ is the gyromagnetic ratio, $\alpha$ is the magnetic damping constant, $\nabla U$ is the energy gradient across a skyrmion, $\bm{F}$ is the equivalent driving force and $\bm{G}=(0,0,G)$ is the gyromagnetic coupling vector where $G=\pm 4\pi$ for a skyrmion. $\mathcal{D}$ is known as the dissipative force tensor.

The first term in Eq.\ (\ref{Thiele}), causes skyrmions to move orthogonally to an applied force, with the direction of movement given by the winding number of the skyrmion. This gyromagnetic coupling term is responsible for the skyrmion Hall effect, where skyrmions move at an angle to a driving current \cite{jiang2017direct}, and it reduces the pinning effectiveness of point defects \cite{Reichhardt2015}. The second term in Eq.\ (\ref{Thiele}) is dissipative in nature, depending on the magnetic damping parameter, $\alpha$, and leads to the skyrmion moving in the direction of the applied force. 

In order to understand the diffusive motion of skyrmions, a thermal force $\bm{F} = \bm{F}_{T}(t)$ \cite{Schutte2014} with $\langle\bm{F}_{T}(t)\rangle = 0$ and non-zero auto-correlation can be added to Eq.\ (\ref{Thiele}). This leads to the tracer diffusion coefficient D of a single skyrmion in a film without grains:

\begin{equation}
\mathsf{D} = k_\mathrm{B} T \frac{\alpha \mathcal{D}}{G^2 + ( \alpha \mathcal{D}) ^2},
\end{equation}
where $k_\mathrm{B}$ is Boltzmann's constant, and $T$ is the temperature. As noted by Sch\"utte \textit{et al}. \cite{Schutte2014}, for a single skyrmion in a clean film the gyromagnetic coupling term tends to dominate --- there is less effective friction for gyratory motion than linear motion. This leads to an inverted dependence on the magnetic damping, that is, skyrmions are expected to diffuse further in films with  higher magnetic damping. 

To go beyond single skyrmion results, we consider a dense skyrmion array. The first additional feature is the repulsive skyrmion-skyrmion interaction potential \cite{zhang2015skyrmion,rozsa2016skyrmions,capic2020skyrmion,brearton2020magnetic} approximately given by:
\begin{equation}
    U_{I}(r) \propto e^{-r/\lambda},
\end{equation}
where $r$ is the distance between skyrmions and $\lambda$ depends on the material parameters and spin profile of the skyrmion. This potential leads to an additional interaction force $F_{I}$ acting on the nearby skyrmions \cite{Pinna2018}, for skyrmions separated by  $\bm{r}$: 
\begin{equation}
     \bm{F_{I}} = -\frac{\mathrm{d}U_{I}}{\mathrm{d} \bm{r}},
\end{equation}
which is also stochastic for skyrmions in dense arrays. Therefore, a stochastic form of the Thiele equation can be written as:
\begin{equation}
     M_{s}{\gamma}^{-1}\bm{G} \times \dot{\bm{C}} + M_{s}{\gamma}^{-1}\alpha \mathcal{D} \cdot \dot{\bm{C}} = \bm{F_{T}} + \bm{F_{I}}. 
\end{equation}
The interaction force can be greater than the thermal force, so that the diffusion coefficient may differ significantly from that predicted by Eq.\ (3). 
\begin{figure}[htb]
    \centering
    \includegraphics[width=1.0\linewidth]{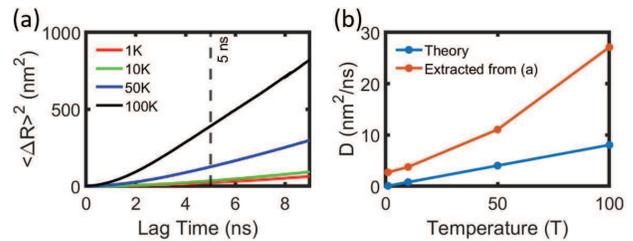}
    \caption{(a) Mean square displacement of skyrmions, $\langle \Delta R \rangle ^2$, in films without grains at different temperatures. The vertical dashed line marks 5 ns, above which $\mathsf{D}$ is extracted from the slopes of the curves. (b) Theoretical value of the skyrmion tracer diffusion coefficient $\mathsf{D}$ compared to the extracted value from (a) in the $5 - 10$ ns range.}
    \label{fig:fig1}
\end{figure}
To quantify this effect, we carry out micromagnetic simulations using the MuMax3 package  \cite{Mumax} including the Dzyaloshinskii-Moriya interaction \cite{Mulkers2017} (DMI) required to create skyrmions in these films. We use parameters similar to those of films which experimentally show a skyrmion phase \cite{YifanTempDMI}, with saturation magnetization $M_s = 1\times10^6$ A/m, exchange stiffness $A= 7 \times 10^{-12}$ J/m, uniaxial anisotropy $K_u = 0.7\times 10^{6}$ J/m$^3$, DMI constant $D = 1.5 \times 10^{-3}$ J/m$^2$, damping parameter $\alpha = 0.3$ and applied magnetic field $B = 0.155$ T. The simulations use a 2 nm $\times$ 2 nm $\times$ 0.8 nm cell with a 512 $\times$ 512 $\times$ 1 grid. We assume the parameters are temperature-independent and limit the maximum temperature in this study to 100 K. At higher temperatures thermally activated skyrmion nucleation and annihilation strongly affect the statistics. Below 100 K the skyrmion lifetime is significantly longer than the simulation time. 

In the simulations, the skyrmions form spontaneously with $\sim$16 nm diameter, giving rise to a dissipative tensor $\mathcal{D} \approx 16$, following Eq.\ (2). The system is allowed to relax for 50 ns in order to reach equilibrium before the data extraction starts, where the equilibrium density of the skyrmions is fixed by the simulation parameters. After relaxation, the skyrmions have a density of $\approx 270/ \mu\textrm{m}^2$, leading to a mean distance between skyrmions of around 65 nm, where the interaction potential is expected to be well-approximated by Eq.\ (4) \citep{brearton2020magnetic}. In order to extract information from the system a sequence of magnetization snapshots of 1000 images with 0.01 ns steps between them are taken and individual skyrmions are tracked through the time steps by the Python library Trackpy \cite{trackpy}. These data are used to extract the temperature dependence of the ensemble-averaged mean square displacement (MSD) $\langle \Delta R \rangle ^2$ as a function of the lag time, $\Delta t$, between images \cite{MetzlerDiffusionReview}. This approach is used because the skyrmions form spontaneously, rather than being created as part of the initialization of the simulation, and so techniques more normally applied to experimental data are appropriate. The results for different simulation temperatures are shown in Fig.\ 1(a) (see also supplementary Video S1 of the simulation at 100 K). The MSD of the skyrmions is approximately linear in time for longer time scales ($>5$ ns) and increases significantly with increasing temperature. The tracer diffusion coefficient $\mathsf{D}$ can be theoretically calculated for a single skyrmion by Eq.\ (3) using parameters from the simulation, as well as extracted from the simulations using:
\begin{equation}\label{Extract_D}
    \langle \Delta R \rangle ^2 = 4\mathsf{D}\Delta t.
\end{equation}
Figure\ 1(b) shows the comparison between the theoretical single skyrmion diffusion coefficients, and values for the tracer diffusion coefficient extracted from the simulations in Fig.\ 1(a). Clearly, for this density of skyrmions, the skyrmion-skyrmion interaction significantly enhances the skyrmion diffusion. Since these simulations are carried out with an equilibrium density of skyrmions, it may be expected that this result will also apply for other material parameters where an array of skyrmions is spontaneously formed. 

Into this picture we now add the effect of grains. Grains arise in sputtered thin films due to the polycrystalline growth of layers. In ultrathin systems that host skyrmions the grain size tends to be of the order of 10 nm \cite{Lee2013}. Differences in the crystal orientation give rise to differing out-of-plane anisotropy in neighbouring grains, with around 10\% variation in the anisotropy strength being typical \cite{JVKim2017}. This means that the grain boundaries act as local anisotropy steps. It has been previously shown that anisotropy steps are capable of producing an effective force on a single skyrmion \cite{Zhou2019}, with implications for devices.
\begin{figure}
    \centering
    \includegraphics[width=1.0\linewidth]{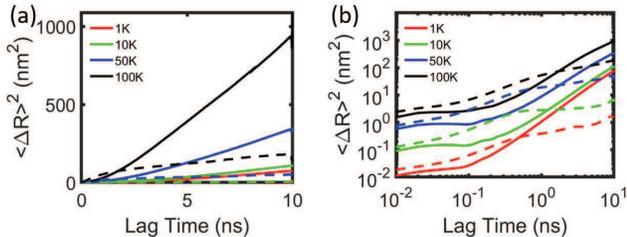}
    \caption{(a) Mean square displacement of skyrmions in films without grains (solid lines) and films with an average grain diameter of 5 nm and an anisotropy variation of 10\% (dashed lines) at different temperatures.  (b) Log-log plot of the data shown in (a).}
    \label{fig:fig2}
\end{figure}
The step in uniaxial anisotropy due to the underlying grains can be incorporated into the Thiele equation as an effective force similar to an energy gradient, by defining the skyrmion energy due to the uniaxial anisotropy, $E_{K_{u}}$. If, for example, the anisotropy varies in the $y$ direction, then: 
\begin{equation}\label{EnergyGradient}
E_{K_{u}}= -K_{u} d  \iint \mathrm{d}x \mathrm{d}y (\bm{m}\cdot \bm{\hat{z}})^2,
 \end{equation}
and the corresponding force:
\begin{equation}\label{GrainForce}
\bm{F_{G}}=-\frac{\mathrm{d}E_{K_{u}}}{\mathrm{d} K_{u}}\frac{\mathrm{d} K_{u}}{\mathrm{d} \bm{y}},
\end{equation}
with $d$ the thickness of the magnetic layer and $\bm{\hat{z}}$ the unit vector in the out-of-plane direction. Assuming only this effective force, the skyrmion velocity described by the Thiele equation can then be separated into ${x}$ and ${y}$ directions given by:
\begin{equation}\label{ThieleInXY}
 \begin{split}
 &x: G \cdot \dot{C_x}-\alpha \mathcal{D} \dot{C_y}=0,
\\&y: G \cdot \dot{C_y}+\alpha \mathcal{D} \dot{C_x}=F_{G}/M_{s}{\gamma}^{-1}.
 \end{split}
\end{equation}
where $C_x$ and $C_y$ give the skyrmion position perpendicular and parallel to the anisotropy variation, respectively. From these equations a drift angle can be derived, which only depends on the material parameters:
\begin{equation}\label{DriftAngle}
\tan\theta=\frac{|\dot{{C_x}}|}{|\dot{{C_y}}|}=\frac{G}{\alpha\mathcal{D}},
\end{equation}
where large $\theta$ means the skyrmion moves nearly perpendicular to the variation in anisotropy, which for a grain boundary would lead to the skyrmion moving along the boundary. Therefore, a skyrmion experiences a net force created by the grain boundaries it overlaps. The strength and direction of the force will depend both on the detail of the grains and the material parameters of the film \cite{Zhou2019}. A complete version of the stochastic Thiele equation for this case can thus be written as:
\begin{equation}
     M_{s}{\gamma}^{-1}\bm{G} \times \dot{\bm{C}} + M_{s}{\gamma}^{-1}\alpha \mathcal{D} \cdot \dot{\bm{C}} = \bm{F_{T}} + \bm{F_{I}} + \bm{F_{G}}. 
\end{equation}

In micromagnetic simulations, the grains are created using a Voronoi pattern \cite{Lel2014,LegrandSub100Motion2017,JVKim2017}, with average grain diameters $l$ from 1 nm to 100 nm, encompassing the skyrmion diameter of $\approx$ 16 nm. The variation of the anisotropy of the grains, $\Delta K_u$, is drawn from a Gaussian distribution with a standard deviation, $\sigma_{K_u}$ which is varied from 1\% to 20\% of the mean anisotropy, $\bar{K_u}$:
\begin{equation}
    \Delta K_u = 100 \times \sigma_{K_u} /\! \bar{K_u},
\end{equation}
where, $\bar{K_u}$ is kept constant at $0.7\times 10^{6}$ J/m$^3$, as for the simulations without grains.
Figure 2 illustrates the dynamics of skyrmion arrays in a film with a grain structure. In Fig.\ 2(a) the data from Fig.\ 1(a), is reproduced (solid lines) with equivalent data obtained in the presence of grains with an average diameter of 5 nm and $\Delta K_u$ of 10\% (dashed lines). The introduction of grains strongly reduces the magnitude of the MSD in the long lag time limit. In Fig.\ 2(b) we show the same data on a log-log plot. This shows that there are in fact two relevant timescales -- at very short lag times up to around 0.1 ns and at lag times above 0.1 ns. Interestingly, in the simulated films with grains, skyrmions have higher mobility than without grains below 0.1 ns. Without grains the skyrmions show signs of gyrotropic motion -- the plateaux up to 0.1 ns -- which is known to reduce the diffusion coefficient \cite{Schutte2014} and which is not obviously present in the films with grains. Here, with 5 nm average grain size, the skyrmions are continuously in contact with the grain boundaries which hinders the gyratory motion induced by the thermal field.

\begin{figure}
    \centering
    \includegraphics[width=1.0\linewidth]{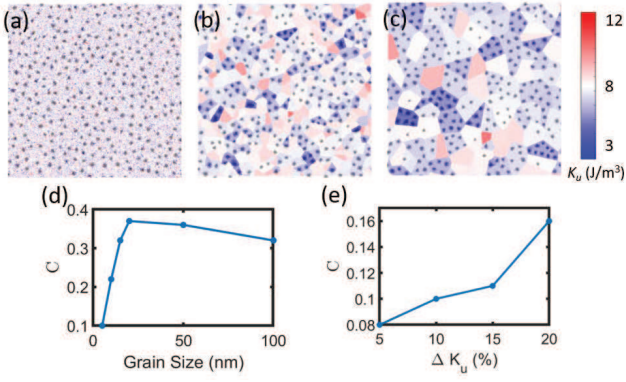}
    \caption{ (a)-(c) Simulation images of skyrmions (black) overlaid on the granular structure encoded by the grain anisotropy using the right-hand scale. The simulation temperature is 100 K and $\Delta K_u = 10 \%$. The average grain size is (a) $5$ nm, (b) $50$ nm and (c) $100$ nm. The images show a simulation of 1024 nm $\times$ 1024 nm. (d) Correlation between skyrmions and $K_u$ as a function of grain size, with $\Delta K_u$ = 10 \%. (e) Correlation between skyrmions and $K_u$ as a function of $\Delta K_u$, with an average grain size of 5 nm.}
    \label{fig:fig3}
\end{figure}

As well as affecting the diffusion of the skyrmions, the presence of grains also affects their spatial distribution. Figures 3(a-c) show simulated magnetization configurations at 100 K, with the skyrmions in black, overlayed on the anisotropy variations for three average grain sizes (see also supplementary videos S2-S4 of the same data). Skyrmions cluster and tend to be on grains with low anisotropy, which is most clearly seen for the simulation with 100 nm grain diameter (Fig.\ 3(c)). Introducing grains also has the effect of slightly lowering the spontaneous skyrmion density, an effect which increases for larger grains and at higher temperatures, so that at 100 K the density reduces by 15 \% for 100 nm average grain size compared to the simulations without grains. However, the clustering for large grains leads to skyrmions having less distance to their nearest neighbour. To quantify the clustering effect, we define a correlation function between the spins in the $z$ direction $S_z$ and the anisotropy $K_u$ in the grains:
\begin{equation}
    C = \frac{\sum_m\sum_n(S^{mn}_z-\bar{S_z})(K^{mn}_u-\bar{K_u})}
    {\sqrt{(\sum_m\sum_n(S^{mn}_z-\bar{S_z})^2)(\sum_m\sum_n(K^{mn}_u-\bar{K_u})^2)}},
\end{equation} 
where $m$ and $n$ are the discrete simulation coordinates in the $x$ and $y$ direction, respectively, $\bar{S_z}$ is the average magnitude of $S_z$, and the sums are taken over all the discrete moments in the simulation. In these simulations, $S_z \approx -1$ in the skyrmion centre, and if this coincides with a site where $K_u < \bar{K_u}$, $C$ is positive. Figure 3(d) shows the extracted spin-anisotropy correlation as a function of grain diameter. The correlation is largest at a grain size of $15$ nm, which corresponds closely to the average skyrmion diameter of $16$ nm. The correlation peak thus occurs when one skyrmion can be confined into one grain. Moreover, Fig.\ 3(e), shows that an increase in $\Delta K_u$ enhances the correlation for a fixed grain size of 5 nm. This means the the pinning of skyrmions by small grains increases with increasing $\Delta K_u$. 
\begin{figure}[bht]
\includegraphics[width=1.0\linewidth]{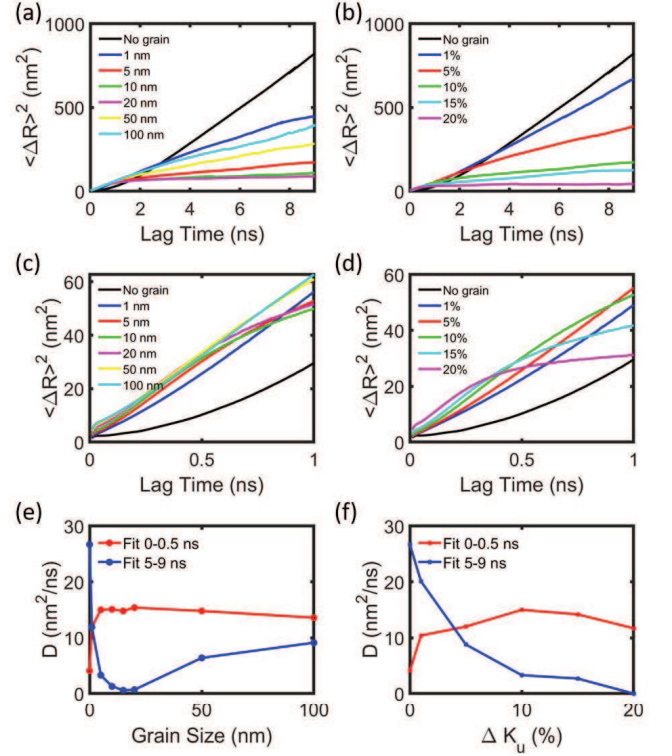}

\caption{(a) Mean square displacement of skyrmions for different grain sizes. $\Delta K_u=10\%$. (b) Mean square displacement of skyrmions for different variations of anisotropy strength. The average grain size is 5 nm. (c) Same data as in (a) for a lag time up to 1 ns. (d) Same data as in (b) for a lag time up to 1 ns. (e) Extracted effective diffusion coefficients $\mathsf{D}$ as a function of grain size. $\Delta K_u = 10 \%$. (f) Extracted effective diffusion coefficients as a function of anisotropy variation. The average grain size is 5 nm. For all simulations the temperature is 100 K.}
\label{fig: fig4}
\end{figure}

In Fig.\ 4 the effect of varying the grain size and anisotropy variation on the MSD of skyrmions is demonstrated. Firstly, in Fig.\ 4(a), the MSD as a function of time for different grain sizes with $\Delta K_u = 10 \%$ is shown, with the same plot in Fig.\ 4(b) for different $\Delta K_u$ for 5 nm grains. Figures 4(c) and 4(d) show the same data as Figs.\ 4(a) and 4(b) respectively, but only plotting the data up to 1 ns lag time, in order to highlight the different regimes seen in the data. In Figs.\ 4(e) and 4(f) we extract the effective diffusion coefficients using Eq. (\ref{Extract_D}) for short ($0-0.5$ ns) and long ($5-9$ ns) lag times.

In Fig.\ 4(e), for the $0-0.5$ ns lag time, $\mathsf{D}$ rapidly increases with grain size for small grains before slowly reducing as the grains size further increases. For $5-9$ ns there is an initial rapid decrease in $\mathsf{D}$ with grain size followed by a slower increase. In Fig.\ 4(f) we find that for long lag times more anisotropy variation leads to a decrease in the diffusion coefficient. For $\Delta K_u =$  20 \% there is no movement of skyrmions at all at longer lag times.  For $0-0.5$ ns, however, $\mathsf{D}$ peaks at intermediate values of anisotropy variation.  

Combining the information of Fig.\ 4 with Eq.\ (12), a detailed explanation for the thermal motion of skyrmions on films with a grain structure can be put forward. Firstly, we discuss the long lag time ($>$5 ns) diffusion coefficients. For grains larger than the skyrmion diameter, the grain boundaries act to prevent the skyrmions from passing to neighbouring grains. The skyrmions are able, however, to diffuse within the grains, leading to an increasing diffusion coefficient with increasing grain size. On approaching a grain boundary from a lower anisotropy side, a skyrmion will experience a force which both pushes it away from the boundary, due to the dissipative term in Eq.\ (10), and moves it along the boundary, due to the gyrotropic term in Eq.\ (10). For a large anisotropy step, the skyrmion needs a large thermal force in order to cross the grain boundary. A skyrmion crossing a grain boundary will also move along the boundary due to the gyrotropic term, with the net direction determined by the ratio of fundamental parameters as given by Eq.\ (11). This sideways motion of a skyrmion crossing a grain boundary has been investigated previously in the case of current-driven skyrmion motion \cite{JVKim2017,Salimath2019,LegrandSub100Motion2017}.
\begin{figure}[htb]
\includegraphics[width=1\linewidth]{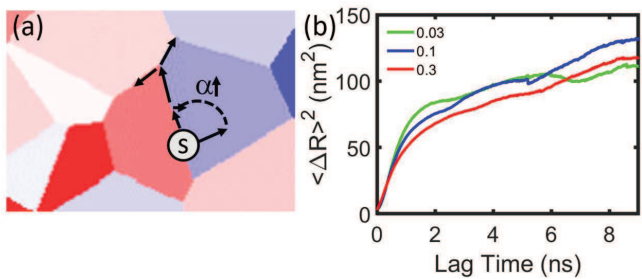}

\caption{(a) Schematic of a skyrmion (S) on a grain boundary. Increasing $\alpha$ will change the direction of skyrmion motion from along the boundary to inside the low anisotropy grain (blue). (b) Mean square displacement of skyrmions as a function of lag time with different damping constant $\alpha$. The average grain diameter is 10 nm and $\Delta K_u= 10\%$, with a simulation temperature of 100 K.
}
\label{fig: fig5}
\end{figure}

In the intermediate case where the grains are of a similar size to the skyrmions, the skyrmion will almost continually interact with the grain boundaries.  Moreover, the skyrmions are effectively pinned by single low anisotropy grains, as demonstrated by the reduction to near zero of the long lag time diffusion coefficient in the case of 15 nm grains. 
For grain sizes significantly smaller than the skyrmion size, the skyrmion is influenced by the sum of the forces from all the grain boundaries it interacts with. This tends to lower the pinning force compared to intermediate grain sizes, however for large $\Delta K_u$, this pinning can still cause $\mathsf{D}$ to become close to zero at long lag times as seen in Fig.\ 4(f) for 5 nm average grain size.  

The effects of grain boundaries at short lag times have a distinct effect to that at long lag times. In Fig.\ 4(c) the effect of variation in the grain diameter is shown at small lag times. The simulation without grains is also included and shows a quadratically increasing MSD with time as expecting for very short lag times from ballistic motion of effectively non-interacting skyrmions. Introducing 1 nm grains causes a similar quadratic curve but with a significantly enhanced diffusivity. Using 5 nm grains causes a further enhancement at very short times, but then the curve flattens at less than 1 ns, with the same features seen for films with 10 nm and 20 nm grains. 

As mentioned in relation to Fig. 2(b), grain boundaries impede thermally-induced gyrotropic motion at very small lag times, under 0.1 ns, which effectively enhances the skyrmion diffusivity. As show in Figs.\ 4(c) and 4(d), the presence of grain boundaries also enhances the diffusivity at longer lag times, on the order of 1 ns. This suggests that the gyrotropic motion induced by the interaction of skyrmions with grain boundaries leads to an increased diffusivity at this intermediate timescale, consistent with findings of an enhanced mobility for current driven motion in the presence of grains \cite{Gong2020}. Considering the effect of a single boundary on a skyrmion, the skyrmion will tend to move along this boundary due to its gyrotropic motion until it meets a vertex. This will lead to a force on the skyrmion which is correlated on a timescale roughly equal to the time it takes the skyrmion to move along a boundary. This force will therefore be correlated on intermediate timescales $(0.1 -1$ ns$)$ and since the motion is largely transverse to the boundary it will act to enhance the diffusivity over this timescale. In Fig.\ 4(f) the effect of changing $\Delta K_u$ at short times is shown. Consistent with the above explanation, higher $\Delta K_u$ leads to higher effective diffusivity at short times, but this is followed by a slight reduction for the highest values of $\Delta K_u$. This dip is associated with the strong reduction of the effective diffusion coefficient seen for $5-9$ ns. Large $\Delta K_u$ leads to very strong pinning at longer times as skyrmions find local minima in $K_u$. This effect extends to shorter times with larger $\Delta K_u$, so although the simulation with $\Delta K_u = 20\%$ has the highest diffusivity up to 0.2 ns, by 0.5 ns it already has entered the strong pinning regime seen at longer lag times.  

The vertices between grains are not expected to provide further pinning separate to the grains. The anisotropy steps between any three grains adjoining a vertex will give gyrotropic skyrmion motion towards or away from the vertex, but all three steps cannot produce motion towards the vertex. This is shown schematically for one vertex in Fig.\ 5(a).  

To confirm that the enhancement of the diffusion coefficients seen at timescales on the order or 1 ns is due to the gyrotropic motion of skyrmions caused by the interaction of the skyrmions with the grain boundaries, we exploit the difference in behaviour due to grain boundaries and the expected pinning free diffusion of skyrmions. For grain-free films the diffusivity is decreased by reducing $\alpha$ as given by Eq.\ (3). 
However, for films with grains, reducing $\alpha$ means the skyrmion will tend to travel more parallel to the grain boundary, as given by Eq.\ (11). Reducing $\alpha$ therefore should lead to an enhancement of the short lag time diffusion whilst causing the long lag time diffusivity to reduce. This is shown in Fig.\ 5(b) where enhancement of the diffusion around 1 ns is clearly shown with reducing $\alpha$ in films with a grain structure. For longer lag times the effect is not so clear. The effect of the grains means that Eq.\ (3) is not correct at long lag times due to pinning by the grains. 
Larger $\alpha$ will tend to drive the skyrmions into low anisotropy grains where they are more likely to be pinned, as shown schematically in Fig.\ 5(a). In this case, therefore, the increase in diffusivity with increased $\alpha$ does not necessarily occur and will depend on the details of the system. The enhancement of diffusivity around 1 ns with reduced $\alpha$ in films with a grain structure is, however, clearly demonstrated.

In conclusion, we have studied in detail the thermal motion of skyrmions in granular films. We find that for large grains, clustering of the skyrmions is induced. The effect of the grains is to enhance the effective diffusivity of skyrmions at short lag times, due to the gyrotropic motion of the skyrmions induced by the grain boundaries. This effect can be understood using the Thiele equation and assuming the grain boundaries have the form of an anisotropy step. For longer times grain boundaries act to strongly reduce the diffusivity, with the effect being most pronounced for grains of a similar size to the skyrmions. These results can be useful to experiments which aim to use the thermal motion of skyrmions to create logic devices, and guide grain engineering as a control factor for these devices.

\begin{acknowledgments}
This work was supported by the Academy of Finland (Grant  Nos.  295269,  306978 and 327804), as well as by the Academy of Finland Centre of Excellence program Quantum Technology Finland (project 312298). We acknowledge the provision of computational resources by the Aalto Science-IT project. We thank See-Chen Ying, Ken Elder and Enzo Granato for useful discussions.
\end{acknowledgments}

\bibliographystyle{aipnum4-1}
\bibliography{ref}

\end{document}